%% file: effective_rixs.tex
\documentclass[aps,prl,twocolumn,showpacs,amsmath,amssymb,superscriptaddress,floatfix]{revtex4-1}

\usepackage{graphicx}
\usepackage{dcolumn}
\usepackage{bm}
\usepackage{amssymb}
\usepackage{booktabs}
\usepackage{color}
\usepackage{microtype,lmodern}
\usepackage[T1]{fontenc}

\usepackage[section]{placeins}

\usepackage{sidecap}

\renewcommand{\tablename}{Table}
\makeatletter\renewcommand{\fnum@table}[1]{\tablename~\thetable.}\makeatother

\DeclareSymbolFont{bbold}{U}{bbold}{m}{n}
\DeclareSymbolFontAlphabet{\mathbbm}{bbold}

\renewcommand{\Im}{\operatorname{Im}}
\newcommand{\ii}{\mathrm{i}}
\newcommand{\llangle}{\langle\!\langle}
\newcommand{\rrangle}{\rangle\!\rangle}
\newcommand{\updownarrows}{\mathbin\uparrow\hspace{-.3em}\downarrow}
\newcommand{\vb}[1]{\mathbf{#1}}

\newcommand{\bra}[1]{\langle#1\rvert}
\newcommand{\ket}[1]{\lvert#1\rangle}
\newcommand{\braket}[2]{\langle #1 | #2 \rangle}
\newcommand{\ketbra}[2]{\lvert #1 \rangle \! \langle #2 \rvert}
\newcommand{\ev}[1]{\langle #1 \rangle}
\newcommand{\mel}[3]{\bra{#1} #2 \ket{#3}}

\usepackage[colorlinks,plainpages=false,linkcolor=blue,urlcolor=blue,citecolor=blue,pdfpagemode=UseNone,pdfstartview=FitBH]{hyperref}

\begin{document}

\title{Non-perturbative series expansion of Green's functions: The Anatomy of \\ Resonant Inelastic X-Ray Scattering in Doped Hubbard Model}

\author{Yi~Lu}
\affiliation{Max-Planck-Institut f\"ur Festk\"orperforschung, Heisenbergstrasse~1, 70569 Stuttgart, Germany}
\affiliation{Institut f\"ur Theoretische Physik, Ruprecht-Karls-Universit\"at Heidelberg, Philosophenweg 19, 69120 Heidelberg, Germany}

\author{Maurits~W.~Haverkort}
\affiliation{Institut f\"ur Theoretische Physik, Ruprecht-Karls-Universit\"at Heidelberg, Philosophenweg 19, 69120 Heidelberg, Germany}

\date{\today}

\begin{abstract}
We present a non-perturbative, divergence-free series expansion of Green's functions using effective operators. The method is especially suited for computing correlators of complex operators as a series of correlation functions of simpler forms. We apply the method to study low-energy excitations in resonant inelastic x-ray scattering (RIXS) in doped one- and two-dimensional single-band Hubbard models. The RIXS operator is expanded into polynomials of spin, density, and current operators weighted by fundamental x-ray spectral functions. These operators couple to different polarization channels resulting in simple selection rules. The incident photon energy dependent coefficients help to pinpoint main RIXS contributions from different degrees of freedom. We show in particular that, with parameters pertaining to cuprate superconductors, local spin excitation dominates the RIXS spectral weight over a wide doping range in the cross-polarization channel.
\end{abstract}

\maketitle

Our understanding in many aspects of physics has benefited tremendously from perturbation theory.
One of its triumphs lies in the practical calculation of various scattering cross sections and correlation functions in many branches of physics ranging from quantum electrodynamics to condensed matter. Although widely successful, such perturbative approaches require the ``perturbing part'' to be small to ensure convergence. This is not always a given, especially for systems that involve different degrees of freedom with comparable energy scales. The divergence of series expansions in quantum electrodynamics has been discussed early on~\cite{Dyson1952}. Fuelled by recent advances in numerical diagrammatic methods aimed to tackle the many-body problem in regimes with competing interactions~\cite{Toschi2007,Rubtsov2008,Kozik2010,Rohringer2013,Ayral2015,Gull2011}, it has been realized that divergent series need to be treated with care~\cite{Gull2011,Lani2012,Schaefer2013,Janis2014,Kozik2015,Stan2015,Ribic2016,Rossi2016,Tarantino2017,Gunnarsson2017}. We here present an alternative expansion scheme for the calculation of correlation functions that does not depend on small parameters and converges for any Hamiltonian. The difference between our projection method and standard perturbation theory is much like the difference between a Fourier series that can describe functions with poles and a Taylor series with finite range of convergence~\cite{SupMat}.

We apply our theory to describe the transition process of RIXS, which has emerged in recent years as a versatile tool for studying the energy-momentum structure of charge, spin, orbital, and lattice excitations in solids~\cite{Ament2011}. RIXS is a second-order process, described by a four-point two-particle Green's function. In the first step, a core electron is promoted into an empty valence state by absorbing a high-energy photon. The created core hole interacts with its environment, and eventually decays into an excited state after certain lifetime by emitting a photon. Detailed information of various elementary excitations is then encoded in the change of energy, momentum, and polarization between the incident and scattered photons. The high photon flux provided by modern synchrotron x-ray sources and its energy-momentum window make RIXS an appealing complement to more established methods such as inelastic neutron scattering (INS). There are clear parallels between the two---both processes can be cast into the form of a momentum space correlation function
\begin{equation}\label{eq:cor}
  \chi(\bm{q},\omega) = -\ii \int  \mathrm{d}t\, e^{\ii \omega t} \ev{O_{\bm{q}}^\dag(t) O_{\bm{q}}(0)}.
\end{equation}
While the results of INS can be directly communicated as the dynamic spin correlation function by identifying the operator $O_{\bm{q}}$ with the spin operator $S_{\bm{q}}$, the complex structure of the RIXS operator originated from its non-trivial intermediate state prevents a direct interpretation of the measured spectra.

One recognizes the challenges in finding an effective theory for RIXS when realizing that the intermediate state in RIXS is the final state in x-ray absorption spectroscopy (XAS). For XAS, it is well known that core-valence interactions lead to excitons or resonances with asymmetric line-shapes, e.g. edge singularities~\cite{Anderson1967,Zaanen1985,Benjamin2013}, combined with multiplets and charge-transfer shake-up excitations~\cite{Haverkort2014,deGroot2008}. No small parameters are present in the intermediate state, and previously proposed series expansions based on the assumption that the core-hole lifetime $1/\Gamma$ is small~\cite{Luo1993,vdB2006,Ament2007} turned out to be non-convergent~\cite{Jia2016,Ament2010}. The lack of concrete understanding of RIXS cross section greatly limits its potential as a true alternative to INS. One acute example is the ongoing debate on the nature of the observed dispersing low-energy RIXS feature in overdoped cuprates~\cite{LeTacon2011,LeTacon2013,Dean2013,Ishii2014,Lee2014,Minola2015,Huang2016,Minola2017,Jia2014,Benjamin2014,KanaszNagy2016}. On one hand, it was interpreted as damped collective magnetic excitations~\cite{LeTacon2011,LeTacon2013,Dean2013,Ishii2014,Lee2014,Minola2015,Minola2017,Jia2014,Huang2016}, supported empirically by its similar momentum-space structure to that of dynamic spin structure factor~\cite{LeTacon2011,Jia2014}. On the other hand, it was argued to be originated from incoherent particle-hole excitations~\cite{Benjamin2014,KanaszNagy2016}, as canonical Fermi-liquid behavior is expected at high doping levels.

In this paper, we present an alternative expansion method that does not depend on small parameters and is free from divergences.
The RIXS operator is expanded into a series of effective operators composed of spin, density, and current operators in the proximity of the core-hole site. Good agreement with the exact solution can be reached within the first few orders, whereby the exact rate of convergence is model and material dependent. This approach provides an unbiased survey of all possible low-energy excitations that couple to the RIXS process with explicit considerations of light polarizations and incident photon energy. The result is an exact mapping of the scattering cross section to a set of intrinsic correlation functions. Such a procedure is generally applicable for a wide class of models.

Following Ref.~\cite{Haverkort2010}, we start by rewriting Eq.~\eqref{eq:cor}, whose imaginary part gives the RIXS intensity as
\begin{equation}\label{eq:cs}
I(\omega_i,\bm{q},\omega) = -\frac{1}{\pi}\Im \bra{0} {R^{\bm{\epsilon}_i\bm{\epsilon}_o}_{\omega_i,\bm{q}}}^\dag \frac{1}{\omega+E_0-H+\ii 0^+} R^{\bm{\epsilon}_i\bm{\epsilon}_o}_{\omega_i,\bm{q}} \ket{0},
\end{equation}
where $\ket{0}$ denotes the initial state of the system with energy $E_0$ determined by the Hamiltonian $H$. $\bm{q}=\bm{k}_i-\bm{k}_o$ and $\omega=\omega_i-\omega_f$ are the momentum and energy transfer between the incoming ($i$) and outgoing ($o$) photons with polarizations $\bm{\epsilon}_i$ and $\bm{\epsilon}_o$, respectively. The $\bm{q}$-dependent RIXS operator $R^{\bm{\epsilon}_i\bm{\epsilon}_o}_{\omega_i,\bm{q}}$ is obtained by a Fourier transform $R^{\bm{\epsilon}_i\bm{\epsilon}_o}_{\omega_i,\bm{q}}=\sum_j e^{\ii\bm{q}\cdot\bm{r}_j}R^{\bm{\epsilon}_i\bm{\epsilon}_o}_{\omega_i,j}$ of the local operator
\begin{equation}\label{eq:op}
R^{\bm{\epsilon}_i\bm{\epsilon}_o}_{\omega_i,j}={T^{\bm{\epsilon}_o}_j}^{\dag}\frac{1}{\omega_i-H+\ii\Gamma}T^{\bm{\epsilon}_i}_j,
\end{equation}
which describes the RIXS process where a core hole is created at site $j$ by a dipole transition operator $T^{\bm{\epsilon}_i}_j$ and subsequently annihilated locally by ${T^{\bm{\epsilon}_o}_j}^\dag$ after some lifetime $1/\Gamma$. A generally valid assumption is imposed here that the core hole is non-dispersive, as the overlap of core-electron wave-functions is usually negligible between different sites.

To understand what excitations can be probed by RIXS, it is favorable to express $R^{\bm{\epsilon}_i\bm{\epsilon}_o}_{\omega_i,j}$ as a sum of effective operators that will consequently turn Eq.~\eqref{eq:cs} into a series of simpler correlation functions as
\begin{equation}\label{eq:genexp}
  R^{\bm{\epsilon}_i\bm{\epsilon}_o}_{\omega_i,j} = \sum_{m,n} \alpha_{m,n}(\bm{\epsilon}_i,\bm{\epsilon}_o,\omega_i) O^{m,n},
\end{equation}
where the operator $O^{m,n}=\ketbra{m}{n}$ brings a state $\ket{n}$ to $\ket{m}$. In cases where $\ket{m}$ and $\ket{n}$ are states in a complete orthonormal Hilbert space, the expansion is trivial and $\alpha_{m,n}$ is readily given as $\alpha_{m,n}= \mel{m}{R}{n}$~\cite{SupMat}. Expansion over a complete basis set is not practical due to the exponentially large size of the Hilbert space. We want to expand $R$ on operators that couple exponentially many states sharing certain property $n$ to exponentially many states having property $m$ in common. Two sufficient conditions for obtaining the expansion coefficients in Eq.~\eqref{eq:genexp} for such operators are:
\begin{equation}
  \begin{split}\label{eq:conditions}
  {O^{m,n}}^{\dag} &= {O^{n,m}}^{\phantom{\dag}}, \\
  O^{s,t} O^{m,n} &= \delta_{t,m} O^{s,n}.
  \end{split}
\end{equation}
Using these relations we obtain $\alpha_{m,n}$ for any given wave function $\ket{\psi}$ by multiplying Eq.~\eqref{eq:genexp} on the right by $O^{u,v} \ket{\psi}$ and left by $ \bra{\psi} O^{t,s}$,
\begin{equation}
  \begin{split}
    \mel{\psi}{O^{t,s} R O^{u,v}}{\psi} &= \sum_{m,n} \alpha_{m,n} \mel{\psi}{O^{t,s}  O^{m,n} O^{u,v}}{\psi}\\
    &= \alpha_{s,u} \mel {\psi}{O^{t,v}}{\psi},
  \end{split}
\end{equation}
from which it follows that
\begin{equation}
\alpha_{s,u}(\bm{\epsilon}_i,\bm{\epsilon}_o,\omega_i) = \frac{ \mel{\psi}{O^{s,s} R^{\bm{\epsilon}_i\bm{\epsilon}_o}_{\omega_i,j} O^{u,u}}{\psi} }{ \mel{\psi}{O^{s,u}}{\psi} },
\end{equation}
where we used the freedom to choose $t=s$, and $v=u$.

For RIXS (or any core-level spectroscopy in general), one should bear in mind its \emph{local} nature defined by the immobile core hole and its short lifetime, which means that the expansion of $R^{\bm{\epsilon}_i\bm{\epsilon}_o}_{\omega_i,j}$ will have a good convergence rate by expanding over operators that act in the proximity of the core-hole site $j$. Let us define the Fock space of a subsystem $\mathcal{L}_L=\otimes^{l \in [L]} H_l$, where $[L]$ denotes the core-hole site ($L=0$) or its up to $L$-th-nearest neighbors ($L\geq 1$). $H_l$ is the single-site Fock space at site $l$. Operators satisfying Eq.~\eqref{eq:conditions} can then be defined as $O_L^{m,n} \equiv \ketbra{\tilde{m}_L}{\tilde{n}_L}\otimes \mathbbm{1}_{\mathcal{R}}$, where $\ket{\tilde{m}_L}$ and $\ket{\tilde{n}_L}$ are orthonormal basis states of $\mathcal{L}_L$ and $\mathbbm{1}_{\mathcal{R}}$ is the identity operator acting on $\mathcal{R}_L=\otimes^{l \notin [L]} H_l$.

In the following we consider the single-band Hubbard model, although we note that generalization to multi-orbital case is straightforward. A multi-orbital expansion restricted to local spin operators can be found in Ref.~\cite{Haverkort2010}. The single-site Fock space in this case is spanned by states with zero, single and double occupations \{$\ket{\varnothing}$, $\ket{\downarrow}$, $\ket{\uparrow}$, $\ket{\updownarrows}$\}. In the zeroth-order ($L=0$) expansion, there are in total $4\times4=16$ operators $O_0^{m,n}$, among which only $\sum_{n=0}^2 {\binom {2} {n}}^2=6$ particle-number-conserving ones couple to the RIXS process, namely
\begin{align}\label{eq:oploc}
  O^{0,0} &= n^{(0)}, & O^{3,3} &= n^{(2)}, \nonumber \\
  O^{1,1} &= \frac{1}{2}n^{(1)} - S_z, & O^{1,2} &= S_x-\mathrm{i}\,S_y,\\
  O^{2,2} &= \frac{1}{2}n^{(1)} + S_z, & O^{2,1} &= S_x+\mathrm{i}\,S_y, \nonumber
\end{align}
which are the linear combinations of local spin operators $S_x$, $S_y$, $S_z$ and density operators $n^{(0)}=(1-n_\uparrow)(1-n_\downarrow)$, $n^{(1)}=\sum_\sigma n_\sigma(1-n_{\bar{\sigma}})$, and $n^{(2)}=n_\uparrow n_\downarrow$ defining the complete set of on-site excitations. Among the other 10 operators, 8 (2) change the particle number by 1 (2), which can combine with operators on neighboring sites to form number-conserving ones in higher-order expansions. An expansion including two (three) sites generates 256 (4096) operators, and 70 (924) of them are number conserving~\cite{SupMat}. Although this is a rapidly growing series, the expansion converges within the first few orders~\footnote{The Fermi velocity is $v_F\sim 10^6$~m/s in a good metal, thus the distance an electron can travel in the intermediate state is $v_F/\Gamma\sim 10$~\AA, corresponding to only a few lattice spacings. See also Ref.~\cite{Savary2015}}.
In addition, the number of operators can be further reduced by considering the local symmetry and light polarizations. For systems with local inversion symmetry, only $S_z$ has non-zero coefficient in the $L=0$ expansion for the cross-polarization channel ($\bm{\epsilon}_i\cdot\bm{\epsilon}^*_o=0$ and assuming $\bm{\epsilon}_i \times \bm{\epsilon}^*_o\parallel z$), or $n^{(0)}$ and $n^{(1)}$ for the parallel-polarization ($\bm{\epsilon}_i\cdot\bm{\epsilon}^*_o=1$) one.

\begin{figure}[t]
  \includegraphics[width=\columnwidth]{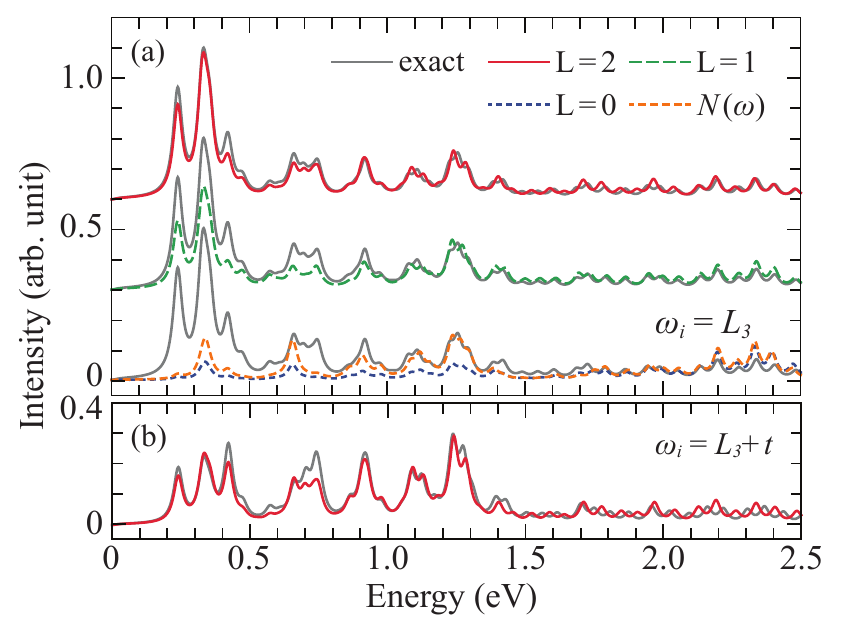}
  \caption{\label{fig:int1dxx} Exact and approximate $\bm{q}$-integrated RIXS cross section for 1D Hubbard model with $n=$ 0.83 in the parallel-polarization channel after subtracting the elastic peak. The incident photon energy is tuned to (a) the $L_3$-resonance and (b) at $t=0.4$~eV higher for the lower panel. The dynamic charge structure factor $N(\omega)$ is also plotted for comparison (see text).}
\end{figure}

The Hamiltonian of the model we study reads
\begin{align}\label{eq:ham}
H = & -t\sum_{\langle i,j\rangle,\sigma}d^\dag_{i,\sigma}d_{j,\sigma}-t^\prime\sum_{\llangle i,j\rrangle,\sigma}d^\dag_{i,\sigma}d_{j,\sigma}+U\sum_i n^d_{i\uparrow}n^d_{i\downarrow}\nonumber \\
& + U_c\sum_{i,\sigma,l,\sigma^\prime}n^d_{i\sigma} n^c_{il\sigma'} + \zeta^c \sum_i \vb{l}^c_i \cdot \vb{s}^c_i,
\end{align}
which is defined on a one-dimensional (1D) chain or a two-dimensional (2D) square lattice. The first line of Eq.~\eqref{eq:ham} describes the hoppings of the ``$d$''-electrons between nearest ($\langle,\rangle$) and next-nearest ($\llangle,\rrangle$) neighboring sites with local Coulomb interaction $U$, and the second line accounts for the repulsion $U_c$ between the $d$ states and the core states ``$c$'' with spin-orbit coupling $\zeta^c$. The numerical results will be evaluated with parameters given in the Supplemental Material~\cite{SupMat}. For both 1D and 2D, the calculation is performed on 12-site clusters with periodic boundary conditions using the many-body package {\sc Quanty}~\cite{Haverkort2016,Lu2014}.

\begin{figure}[t]
  \includegraphics[width=\columnwidth]{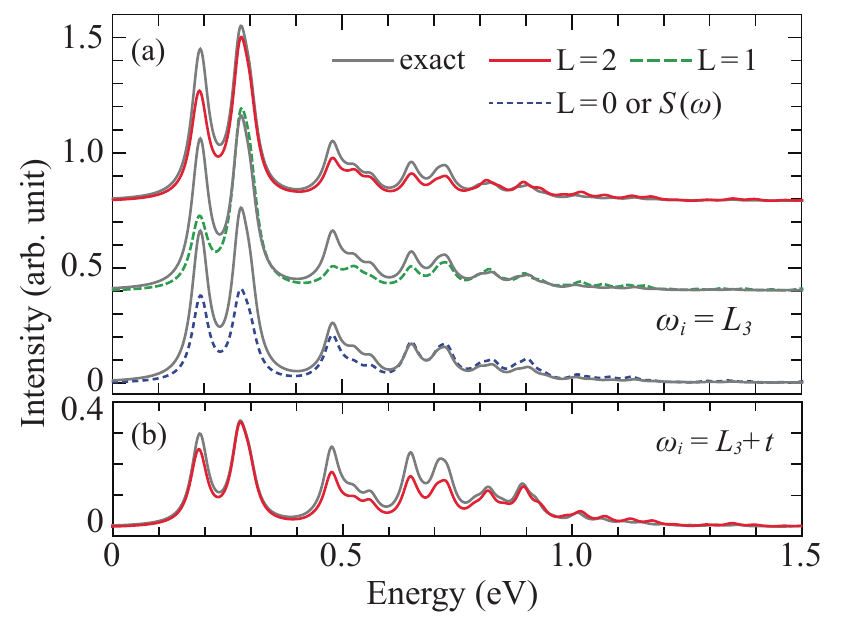}
  \caption{\label{fig:int1dxy} Same as Fig.~\ref{fig:int1dxx}, for the cross-polarization channel. The dynamic spin structure factor $S(\omega)$ is equivalent to the $L=0$ expansion.}
\end{figure}

We first show the results for the 1D case. The exact $\bm{q}$-integrated RIXS cross sections in the parallel-polarization channel for the hole-doped case with occupation $n=0.83$ are plotted in Fig.~\ref{fig:int1dxx}, in comparison to the approximate ones with effective operators $O^{m,n}_L$ expanded up to next-nearest neighbors ($L\leq2$). The incident photon energy $\omega_i$ is tuned to the maximum of the XAS spectra (``resonance'')~\cite{SupMat} for Fig.~\ref{fig:int1dxx}(a) and also ``detuned'' at $t=0.4$ eV higher for Fig.~\ref{fig:int1dxx}(b). All spectra are broadened by a Lorentzian function with full width at half maximum of $0.05t$ for plotting.

Within the $L=0$ expansion, regardless of doping and incident photon energy (see Fig.~\ref{fig:int1dxx} and Supplemental Material~\cite{SupMat}), the approximate cross sections only resemble the exact ones on the higher-energy end, while the low-energy spectral weight is suppressed. This highlights the importance of non-local charge fluctuations as pointed out by earlier studies~\cite{Benjamin2014,Jia2016}. The missing spectral weight can be largely restored by including the neighboring sites in the expansion, and the $L=2$ approximation reproduces almost fully the exact solution. The convergence rate can be quantified by comparing the spectral moments $\mu_n=\int \omega^n A(\omega) d\omega$ between the exact and approximate RIXS cross sections. A good convergence of the spectral weight $\mu_0$ and its center of mass $\mu_1/\mu_0$ is reached for $L\geq 1$ with different $\omega_i$ values both before and after the resonance~\cite{SupMat}.

To address the relation between RIXS and the charge response function, we performed a partial expansion in the $L=0$ subspace using only the density operator $n=n^{(1)}+2n^{(2)}$. The resultant dynamic charge structure factor $N(\omega)$ is shown in Fig.~\ref{fig:int1dxx}(a). The large discrepancy between the RIXS cross section and $N(\omega)$ suggests that a direct association between the two should be discouraged. Their fundamental difference is not difficult to understand. In the presence of large on-site repulsion, the local charge fluctuation is associated with an energy scale of $U$ and thus strongly suppressed. The RIXS process, on the other hand, involves other low-energy excitations as shown by the additional effective operators such as $\vb{S}_0 \cdot \vb{S}_1$~\footnote{The subscripts are the site indices, where 0 denotes the core-hole site, and $l=1,2,\dots$ the bonding combination of the neighborings sites with $l$ lattice spacings to the core-hole site.}, which corresponds to a two-spin excitation and contributes to the lower-energy RIXS spectral weight.

\begin{figure}[t]
  \includegraphics[width=0.9\columnwidth]{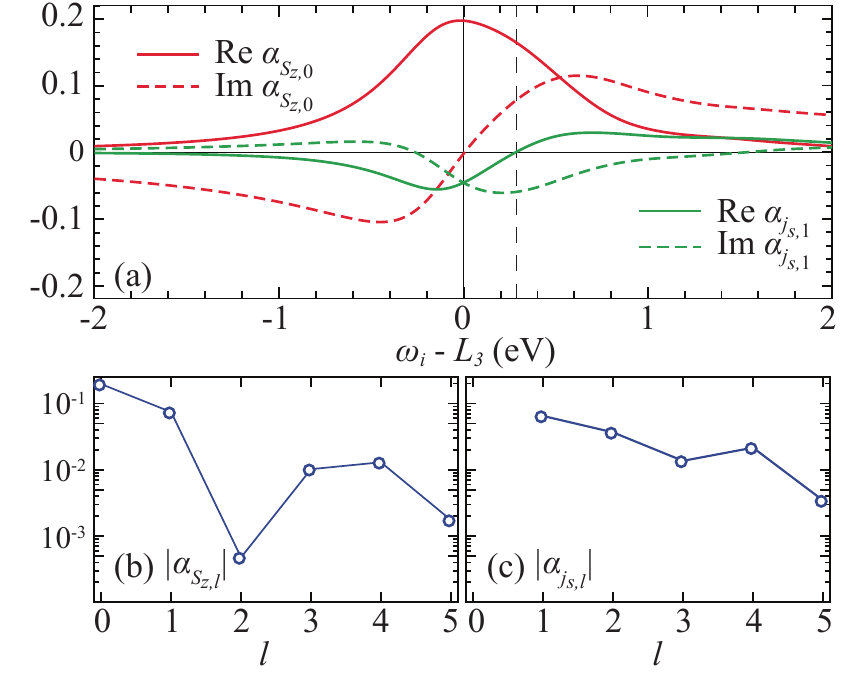}
  \caption{\label{fig:coeff} (a) Real and imaginary parts of $\alpha_{S_{z,0}}(\omega_i)$ and $\alpha_{j_{s,1}}(\omega_i)$ with $n=0.83$. Their resonant energies determined by the maximum of $|\alpha|$ are marked by solid and dashed vertical lines, respectively. (b) $|\alpha_{S_{z,l}}|$ and (c) $|\alpha_{j_{s,l}}|$ evaluated at $\omega_i=L_3$ as a function of $l$.}
\end{figure}

Figure~\ref{fig:int1dxy} shows the 1D exact and approximate RIXS cross sections in the cross-polarization channel. Similar to the parallel-polarization results, an overall improvement of the approximation can be achieved by extending the expansion range. Contrary to the former case where the inclusion of non-local excitation is essential, the majority of the spectral weight is already captured here at all dopings by a local expansion~\cite{SupMat}. This observation indicates that the RIXS spectrum consists mainly of dynamic spin structure factor not only in undoped magnetic insulators~\cite{Haverkort2010} but also up to a high doping range. This seemingly surprising result can be explained by examining the relevant non-local effective operators. The cross-polarization channel couples to those operators that break inversion symmetry such as a spin-pair exchange $j_{s,l}=(c_{0,\sigma}^\dag c_{l,\sigma}-c_{0,\bar \sigma}^\dag c_{l,\bar \sigma}) - \mathrm{h.c.}$ between the core-hole site and the $l$-th-nearest-neighbor sites. Such processes are strongly prohibited compared to a local spin excitation when large local repulsion $U$ is present. Nonetheless, these non-local contributions can give rise to a nontrivial $\omega_i$ dependence of the RIXS spectra. While the energy of local spin excitations is $\omega_i$ independent, as shown by the constant $\mu_1/\mu_0$ values of $L=0$ spectra [or equivalently $S(\omega)$]~\cite{SupMat}, the non-local spin-flip process will move to higher energy with increasing $\omega_i$. The difference of $\omega_i$ dependence is crucial for pinpointing the nature of the experimentally observed low-energy feature of the cross-polarization RIXS in doped cuprates~\cite{Benjamin2014,Minola2015,Huang2016,Minola2017}.

Figure~\ref{fig:coeff}(a) shows the coefficient $\alpha(\omega_i)$ for $S_{z,0}$ and $j_{s,1}$ for $n=0.83$.
While the resonant energy of $\alpha_{S_{z,0}}$ coincides with the XAS, that of $\alpha_{j_{s,1}}$ is located at $\sim 0.3$~eV higher. The relative weight of non-local excitations thus increases by detuning the incident photon to higher energies.
Therefore, while the dominant spectral weight in the cross-polarization channel still originates from the local spin excitations at moderate doping levels (Fig.~\ref{fig:int1dxy}), the relative increase of non-local contributions at high detuning energies may give rise to a fluorescence-like behavior~\cite{SupMat}. Figs.~\ref{fig:coeff}(b) and \ref{fig:coeff}(c) show the expansion coefficients of operators $S_{z,l}$ and $j_{s,l}$ that appear in higher-order expansions. The near exponential convergence rate as a function of lattice spacing $l$ confirms the assumption that the most important excitations happen locally around the core-hole site.

\begin{figure}[t]
  \includegraphics[width=\columnwidth]{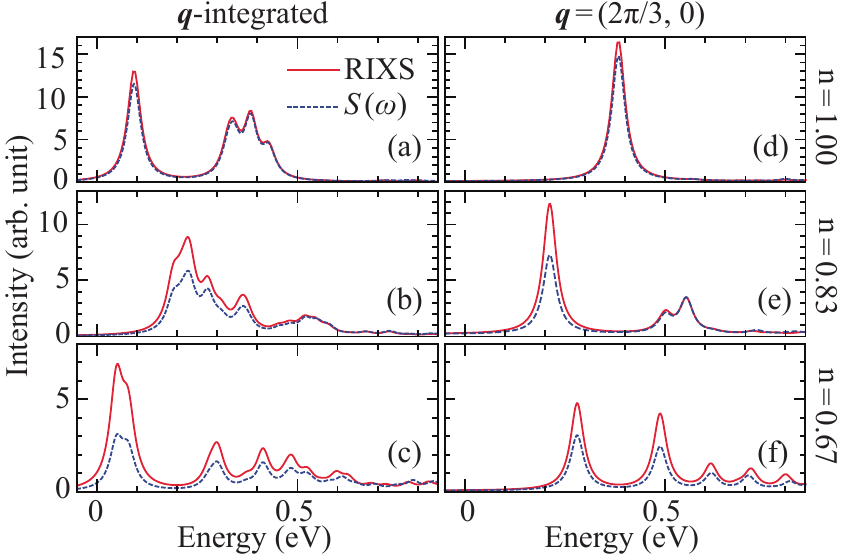}
  \caption{\label{fig:int2dxy} (a)-(c) $\bm{q}$-integrated and (d)-(f) $\bm{q}=(2\pi/3, 0)$ cross-polarization RIXS cross section (solid) and $S(\omega)$ (dashed) for 2D Hubbard model. The incident photon energy is tuned to $L_3$-resonance. Top, middle, and bottom panels show results for $n=$ 1.00, 0.83, and 0.67, respectively.}
\end{figure}

In the last part we address the relation between the cross-polarization RIXS and $S(\omega)$ in hole-doped 2D Hubbard model using the presented method. Figure~\ref{fig:int2dxy} shows the $\bm{q}$-integrated and $\bm{q}=(2\pi/3,0)$ RIXS cross sections together with $S(\omega)$ obtained by $L=0$ expansion. The higher order expansions are shown in the supplemental material~\cite{SupMat}. We find that the convergence rate for the 2D Hubbard model is slightly slower than for the 1D case. At half filling, as shown in Fig.~\ref{fig:int2dxy}(a) and (b), the cross-polarised RIXS probes nearly exact the local spin excitations at low energy, as evidenced by their almost identical spectral weight and line shape. Upon hole doping, while the discrepancy between the two increases due to growing non-local charge and spin fluctuations, the RIXS spectral weight originated from local excitations remains dominant. Even for highly overdoped $n=0.67$ case [Fig.~\ref{fig:int2dxy}(c) and (f)], which was deemed to be fully describable by Fermi liquid theory~\cite{Benjamin2014}, the local excitations still constitute about 60 percent of the total spectral weight, in line with recent experiments~\cite{Minola2017}. This finding supports earlier ``paramagnon'' interpretations~\cite{LeTacon2011,LeTacon2013,Minola2015} of the low-energy feature in cross-polarization RIXS on doped cuprates and suggests that considerations based purely on quasiparticles cannot account for the full spectral weight.

In summary, we presented a method to generate a divergence-free series expansion of Green's functions using effective operators. We applied the method to expand the RIXS operator into a sum of spin, density, and current operators. The result is an exact mapping of the scattering cross section to a set of intrinsic correlation functions, independent of the model and parameters used. The coefficients of the effective operators encode the energy and polarization dependence of RIXS and help to identify the main excitations contributing to the RIXS spectral weight. A quantitative connection between the RIXS cross section and intrinsic correlation functions is provided. Using realistic models tailored to specific materials may help to resolve confusions in the understanding of current measurements and guide future experimental works.

We thank L.\,J.\,P. Ament, T.\,P. Devereaux and K. Wohlfeld for discussion.

\clearpage
\clearpage
\clearpage
\setcounter{page}{1}
\input{supplemental_material.tex}

\end{document}

%% file: supplemental_material.tex
\clearpage
\appendix
\onecolumngrid
\begin{center}
\textbf{\large{Supplemental Material for \\ ``Non-perturbative series expansion of Green's functions: The Anatomy of \\ Resonant Inelastic X-Ray Scattering in Doped Hubbard Model''}}
\end{center}

\setcounter{figure}{0}
\setcounter{table}{0}
\setcounter{equation}{0}
\renewcommand{\thefigure}{S\arabic{figure}}
\renewcommand{\thetable}{S\Roman{table}}
\renewcommand{\theequation}{S\arabic{equation}}

\section{Convergence of series expansions}

Perturbative solutions of the eigenvalues, eigenvectors, or Green's functions for a given Hamiltonian $H$ written as the sum of a bare Hamiltonian $H_0$ and perturbing Hamiltonian $H_1$ can be written as a power series in the perturbing Hamiltonian with expansion coefficients given as derivatives of $H$ with respect to $H_1$. This is a feature perturbative methods have in common with a Taylor series expansion of a function. The correspondence between a Taylor series and more general perturbative expansions implies that similar convergence criteria apply. For a realistic interacting Hamiltonian, perturbative solutions such as diagrammatic series expansions (e.g. Dyson series) have a tendency to diverge, even with moderate interaction strength. In this letter we provide a non-divergent alternative. Instead of creating a power series in the perturbing interaction, we use a projection method to expand the Green's functions. This is analogous to approximating a function with a Fourier series instead of a Taylor expansion. The convergence criteria of the former are in general more easily satisfied. In particular, poles, discontinuities or divergences in the complex plane of the original function do not pose problems for the Fourier expansion.

In the following two subsections we discuss the convergence of our projection method for operators. In the first subsection we illustrate the impact of a divergent series by comparing the Taylor and Fourier expansions of the function $\frac{1}{x^2+1}$. In the second subsection we prove that the projection method we propose in the main text leads to a convergent series.

\subsection{Taylor v.s. Fourier expansion of functions}

\begin{figure*}[htp]
  \includegraphics[width=\textwidth]{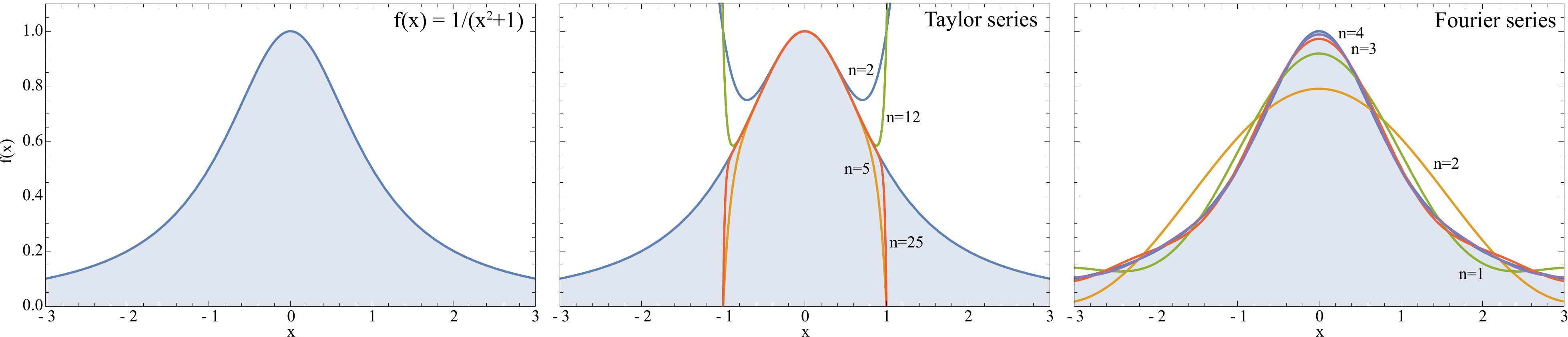}
  \caption{\label{Sfig:TaylorvsFourier} Taylor and Fourier series approximation to the function $f(x)=\frac{1}{x^2+1}$ for different number of expansion coefficients.}
\end{figure*}

To illustrate the difference between a divergent and a convergent series expansion, we compare the Taylor and Fourier expansions of the function
\begin{equation}
f(x)=\frac{1}{x^2+1}.
\end{equation}
The function is shown in the left panel of Fig. \ref{Sfig:TaylorvsFourier}. It is continues and infinitely differentiable on the real axis. Being an even function, only even-order terms arise in its expansion. The Taylor and Fourier series are given as:
\begin{eqnarray}
f(x)&=&\sum_{n=0}^{\infty} \alpha_n^T x^{2n},\\
\nonumber f(x)&=&\sum_{n=0}^{\infty} \alpha_n^F \cos(n x),
\end{eqnarray}
with the expansion coefficients
\begin{equation}
\alpha_n^T = \frac{1}{(2n)!}\frac{d^{2n}}{dx^{2n}} f(x)\big|_{x=0},
\end{equation}
and
\begin{equation}
\alpha_n^F = \frac{1}{(1+\delta_{0,n})\pi}\int_{-\pi}^{\pi}\cos(n x) f(x).
\end{equation}

As $f(x)$ has poles in the complex plane at $x=\pm\mathrm{i}$, the Taylor series only converge for $|x|<1$. This can be seen in the middle panel of Fig. \ref{Sfig:TaylorvsFourier}, where the series is truncated at $n=2,5,12$ and $25$.
The expansion does not approximate well to the original function with increasing $n$ outside of the convergence range $|x|<1$. A Fourier series, on the other hand, nicely converges to the original function, as shown in the right panel of Fig. \ref{Sfig:TaylorvsFourier}.

\subsection{Convergence of the effective operator series}

In Eq. 4 of the main article ($R = \sum_{m,n} \alpha_{m,n} O^{m,n}$), we define an exact mapping of the RIXS operator to a set of effective operators. The operators are defined on a complete set of orthonormal states $\ket{m}$ with $\braket{m}{n}=\delta_{m,n}$ and $\sum_m \ketbra{m}{m}=1$ as $O^{mn}=\ketbra{m}{n}$. The expansion coefficients of operator $R$ expanded on this complete set are given as
\begin{equation}
\alpha_{m,n}= \mel{m}{R}{n}
\end{equation}
which, as stated in the main text can be derived by multiplying Eq. 4 from the left and right with $\bra{m}$ and $\ket{n}$.

Absolute point-wise convergence of this series expansion can be proven by realizing that the RIXS intensity is bounded. In a RIXS process, the RIXS operator promotes a given initial state $\ket{i}$ into $\ket{\psi_R}$ as
\begin{equation}
  \begin{split}
    \ket{\psi_R}=&R\ket{i}\\
    =&\sum_{m,n}\alpha_{m,n}O^{m,n}\ket{i}\\
    =&\sum_{m}\alpha_{m,i}\ket{m}.
  \end{split}
\end{equation}
As the RIXS intensity is bounded by some constant $c$ ($\braket{\psi_R}{\psi_R}\leq c$), it follows that
$\sum_m \left|\alpha_{m,i}\right|^2 \leq c$,
which is a sufficient condition for the series to converge absolutely.

Knowing that these projection methods lead to a convergent series, Eq. 5 to Eq. 8 focus on the question on how to optimize the rate of convergence of this series.

\section{Effective operators for $L=1$ expansion}

Equation 8 shows the 6 effective operators used for the expansion of the RIXS operator locally on the core-hole site ($L=0$). We here provide the 70 operators that arise in the $L=1$ expansion.

There are several ways to create the 70 operators needed for the $L=1$ expansion. One should note that these 70 operators define a (non-unique) basis, and any linear combination of these operators can be used as a basis set to expand the RIXS operator on. This freedom of choice of set of orthonormal operators to project to is much like the difference between a Fourier series and Chebyshev polynomial expansion for functions. In our calculation, we created the non-local operators as products of the complete sets of local operators acting on the core-hole site and its neighboring sites.

As a didactic introduction, we derive here the 70 possible operators following the same arguments as done in the main text for the $L=0$ case. The single-site Fock space containing all states with locally 0, 1 or 2 electrons is spanned by the 4 states \{$\ket{\varnothing}$, $\ket{\downarrow}$, $\ket{\uparrow}$, $\ket{\updownarrows}$\}. The basis set for two sites are given as their outer product, which contains 16 states \{$\ket{\varnothing,\varnothing}$, $\ket{\downarrow,\varnothing}$, $\ket{\uparrow,\varnothing}$, $\ket{\updownarrows,\varnothing},\ket{\varnothing,\downarrow}$, $\ket{\downarrow,\downarrow}$, $\ket{\uparrow,\downarrow}$, $\ket{\updownarrows,\downarrow},\ket{\varnothing,\uparrow}$, $\ket{\downarrow,\uparrow}$, $\ket{\uparrow,\uparrow}$, $\ket{\updownarrows,\uparrow},\ket{\varnothing,\updownarrows}$, $\ket{\downarrow,\updownarrows}$, $\ket{\uparrow,\updownarrows}$, $\ket{\updownarrows,\updownarrows}$\}. There are thus $16\times16=256$ linear independent operators possible within this Fock space. We label the 16 basis states from $\ket{1}=\ket{\varnothing,\varnothing}$ to $\ket{16}=\ket{\updownarrows,\updownarrows}$. Considering the conditions Eq. 5, we define $O^{m,n}=\ketbra{m}{n}$, and thus $O^{1,1}=\ketbra{\varnothing,\varnothing}{\varnothing,\varnothing}$, $\dots$, $O^{1,16}=\ketbra{\varnothing,\varnothing}{\updownarrows,\updownarrows}$, $\dots$, and $O^{16,16}=\ketbra{\updownarrows,\updownarrows}{\updownarrows,\updownarrows}$. As the RIXS process is particle number conserving, the coefficients of operators like $O^{1,16}$ are zero in the expansion, leaving only 70 operators relevant for the expansion. The above procedure defines a valid and straightforward method to obtain all operators for expansion of any order. However, it does not always give a set of operators which provide the most physical insight into the RIXS process. From here on one can however linear combine them to the more familiar operators, such as the identity, spin, number, and current operators.

Table~\ref{Stab:L1coeffs} shows one possible basis set of operators for the $L=1$ expansion. Subscripts 0 and 1 denote the core-hole site and the bonding linear combination of the nearest-neighbor sites, respectively. Here we define the number conserving local operators of each site as the identity $\mathbbm{1}$, number $n$, double occupation $n^{(2)}=n_\uparrow n_\downarrow$, and spin operators $S^\pm$ and $S_z$. This is equivalent to Eq. 8, but gives a simpler form of the multi-site operators. Creation and annihilation operators with superscript such as $c^{(1)}$ is a shorthand for $c\,n^{(1)}$. Note that operators of the form $c^\dag_{0\downarrow}c_{1\downarrow}$ and $c^\dag_{0\uparrow}c_{1\uparrow}$ can be linearly combined as $c^\dag_{0\downarrow}c_{1\downarrow} \pm c^\dag_{0\uparrow}c_{1\uparrow}$, with the $+$ ($-$) combination only coupling to the parallel (cross) polarization channel.

\begin{table*}[htb]
\centering
\begin{ruledtabular}
\begin{tabular}{clclclclcl}
\toprule
\multicolumn{10}{c}{$L=1$} \\ \hline
1& $\mathbbm{1} $ &
2& $n_0 $ &
3& $n^{(2)}_0 $ &
4& $S^+_0 $ &
5& $S^-_0 $ \\
6& $S_{0z} $ &
7& $ n_1$ &
8& $n_0 n_1$ &
9& $n^{(2)}_0 n_1$ &
10& $S^+_0 n_1$ \\
11& $S^-_0 n_1$ &
12& $S_{0z} n_1$ &
13& $ n^{(2)}_1$ &
14& $n_0 n^{(2)}_1$ &
15& $n^{(2)}_0 n^{(2)}_1$ \\
16& $S^+_0 n^{(2)}_1$ &
17& $S^-_0 n^{(2)}_1$ &
18& $S_{0z} n^{(2)}_1$ &
19& $ S^+_1$ &
20& $n_0 S^+_1$ \\
21& $n^{(2)}_0 S^+_1$ &
22& $S^+_0 S^+_1$ &
23& $S^-_0 S^+_1$ &
24& $S_{0z} S^+_1$ &
25& $ S^-_1$ \\
26& $n_0 S^-_1$ &
27& $n^{(2)}_0 S^-_1$ &
28& $S^+_0 S^-_1$ &
29& $S^-_0 S^-_1$ &
30& $S_{0z} S^-_1$ \\
31& $ S_{1z}$ &
32& $n_0 S_{1z}$ &
33& $n^{(2)}_0 S_{1z}$ &
34& $S^+_0 S_{1z}$ &
35& $S^-_0 S_{1z}$ \\
36& $S_{0z} S_{1z}$ &
37& $c^{(1)}_{0\downarrow} {c^{(0)}_{1\downarrow}}^\dag$ &
38& $c^{(1)}_{0\uparrow} {c^{(0)}_{1\downarrow}}^\dag$ &
39& $c^{(2)}_{0\downarrow} {c^{(0)}_{1\downarrow}}^\dag$ &
40& $c^{(2)}_{0\uparrow} {c^{(0)}_{1\downarrow}}^\dag$ \\
41& $c^{(1)}_{0\downarrow} {c^{(0)}_{1\uparrow}}^\dag$ &
42& $c^{(1)}_{0\uparrow} {c^{(0)}_{1\uparrow}}^\dag$ &
43& $c^{(2)}_{0\downarrow} {c^{(0)}_{1\uparrow}}^\dag$ &
44& $c^{(2)}_{0\uparrow} {c^{(0)}_{1\uparrow}}^\dag$ &
45& $c^{(1)}_{0\downarrow} {c^{(1)}_{1\downarrow}}^\dag$ \\
46& $c^{(1)}_{0\uparrow} {c^{(1)}_{1\downarrow}}^\dag$ &
47& $c^{(2)}_{0\downarrow} {c^{(1)}_{1\downarrow}}^\dag$ &
48& $c^{(2)}_{0\uparrow} {c^{(1)}_{1\downarrow}}^\dag$ &
49& $c^{(1)}_{0\downarrow} {c^{(1)}_{1\uparrow}}^\dag$ &
50& $c^{(1)}_{0\uparrow} {c^{(1)}_{1\uparrow}}^\dag$ \\
51& $c^{(2)}_{0\downarrow} {c^{(1)}_{1\uparrow}}^\dag$ &
52& $c^{(2)}_{0\uparrow} {c^{(1)}_{1\uparrow}}^\dag$ &
53& ${c^{(0)}_{0\downarrow}}^\dag c^{(1)}_{1\downarrow}$ &
54& ${c^{(0)}_{0\uparrow}}^\dag c^{(1)}_{1\downarrow}$ &
55& ${c^{(1)}_{0\downarrow}}^\dag c^{(1)}_{1\downarrow}$ \\
56& ${c^{(1)}_{0\uparrow}}^\dag c^{(1)}_{1\downarrow}$ &
57& ${c^{(0)}_{0\downarrow}}^\dag c^{(1)}_{1\uparrow}$ &
58& ${c^{(0)}_{0\uparrow}}^\dag c^{(1)}_{1\uparrow}$ &
59& ${c^{(1)}_{0\downarrow}}^\dag c^{(1)}_{1\uparrow}$ &
60& ${c^{(1)}_{0\uparrow}}^\dag c^{(1)}_{1\uparrow}$ \\
61& ${c^{(0)}_{0\downarrow}}^\dag c^{(2)}_{1\downarrow}$ &
62& ${c^{(0)}_{0\uparrow}}^\dag c^{(2)}_{1\downarrow}$ &
63& ${c^{(1)}_{0\downarrow}}^\dag c^{(2)}_{1\downarrow}$ &
64& ${c^{(1)}_{0\uparrow}}^\dag c^{(2)}_{1\downarrow}$ &
65& ${c^{(0)}_{0\downarrow}}^\dag c^{(2)}_{1\uparrow}$ \\
66& ${c^{(0)}_{0\uparrow}}^\dag c^{(2)}_{1\uparrow}$ &
67& ${c^{(1)}_{0\downarrow}}^\dag c^{(2)}_{1\uparrow}$ &
68& ${c^{(1)}_{0\uparrow}}^\dag c^{(2)}_{1\uparrow}$ &
69& $c_{0\downarrow} c_{0\uparrow} c^\dag_{1\downarrow} c^\dag_{1\uparrow}$ &
70& $c^\dag_{0\downarrow} c^\dag_{0\uparrow} c_{1\downarrow} c_{1\uparrow}$ \\
\bottomrule
\end{tabular}
\end{ruledtabular}
\caption{A possible choice of operators for $L=1$ expansion.}\label{Stab:L1coeffs}
\end{table*}

\begin{figure}[htb]
  \includegraphics[width=0.5\linewidth]{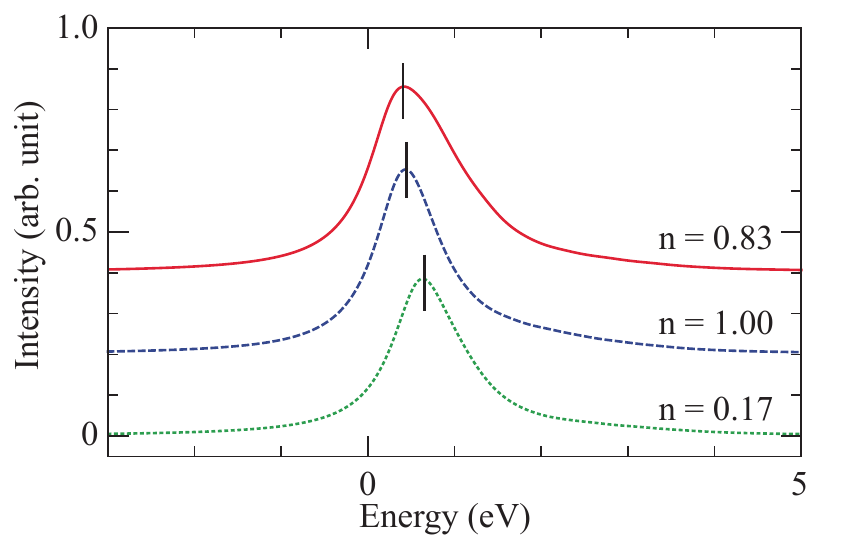}
  \caption{\label{Sfig:xas1d} ``$L_3$'' edge of the XAS spectra of 12-site 1D Hubbard model with average occupation $n$ = 0.83, 1.00, and 1.17. The vertical bars mark the maximum of the intensity. The parameters are listed in the main text.}
\end{figure}

\section{XAS of 1D Hubbard model}

Figure~\ref{Sfig:xas1d} shows the calculated XAS spectra for three different doping levels with occupation $n$ = 0.83 (hole-doped), 1.00 (half-filled), and 1.17 (electron-doped). The parameters are chosen most relevant to the Cu $L$ edge in cuprates, with $t=0.4$~eV, $t^\prime=-0.2t$, $U_c=1.2U$ and $\zeta^c=13.5$~eV. The Coulomb repulsion is defined as $U=2zt$, with $z=2$ the coordination number for 1D and $z=4$ for 2D. This on one hand defines the same magnetic energy scale for both cases with approximately $zJ=2zt^2/U=t$, and on the other hand gives a commonly used value $U=8t$ for 2D. The inverse core-hole life time is chosen as $\Gamma=t$ for the intermediate states.

\section{Overview of exact and approximate RIXS cross sections for 1D Hubbard model}

\begin{figure}[htp]
  \includegraphics[width=0.75\linewidth]{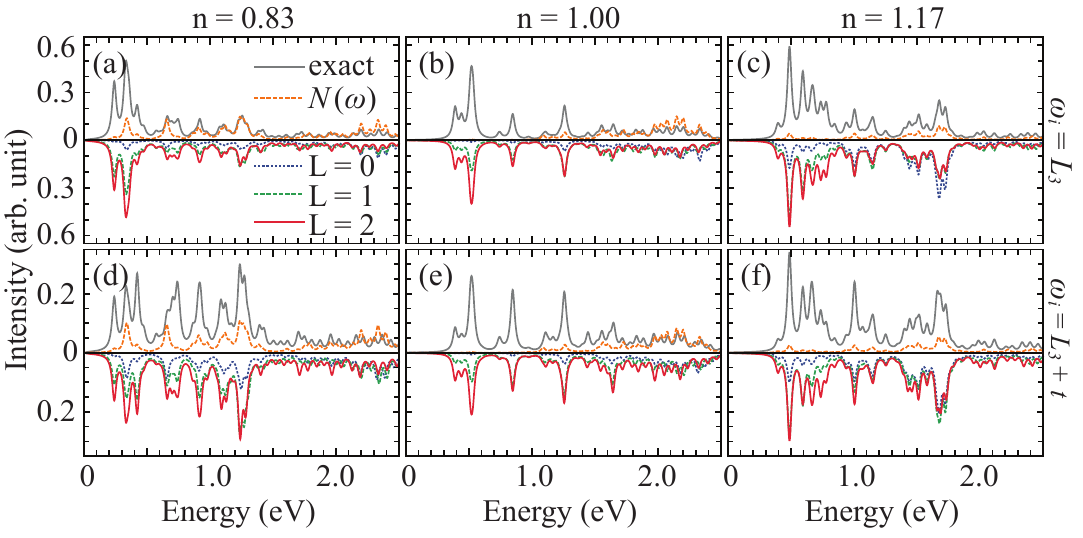}
  \caption{\label{Sfig:int1dxx} (a)-(f) Exact and approximate $\vb{q}$-integrated RIXS cross section for 1D Hubbard model in the parallel-polarization channel after subtracting the elastic peak. The incident photon energy is tuned to $L_3$-resonance for (a)-(c) and at $t=0.4$~eV higher for (e)-(f). Left, middle, and right panels show results for $n=$ 0.83, 1.00, and 1.17, respectively. The dynamic charge structure factor $N(\omega)$ is also plotted for comparison (see main text).}
\end{figure}

\begin{figure}[htp]
  \includegraphics[width=0.75\linewidth]{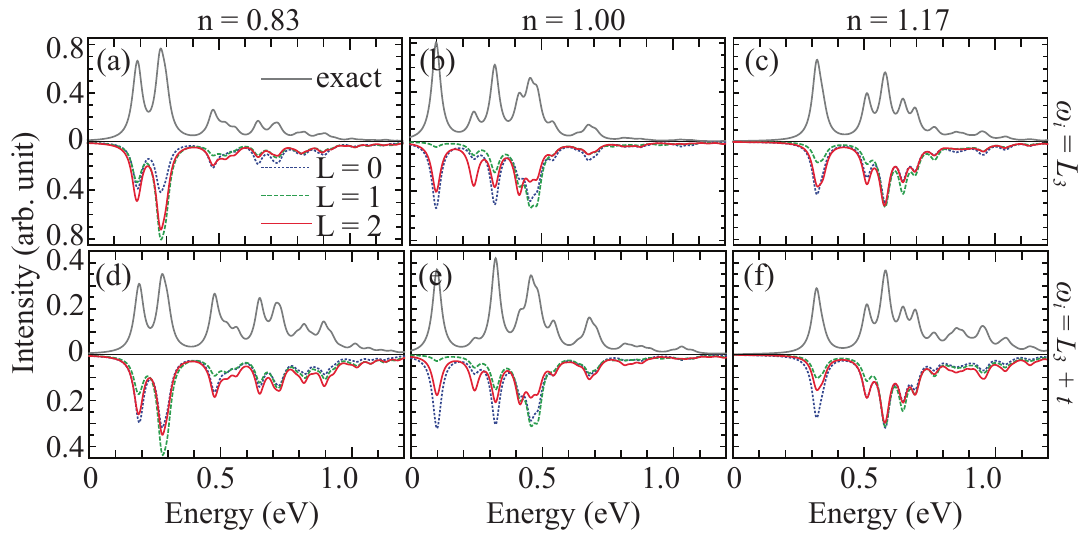}
  \caption{\label{Sfig:int1dxy} Same as Fig.~\ref{Sfig:int1dxx}, for the cross-polarization channel. The dynamic spin structure factor $S(\omega)$ is equivalent to the $L=0$ expansion.}
\end{figure}

\clearpage
\section{Zeroth and first moment of the exact and approximate RIXS cross sections for 1D Hubbard model}

\begin{figure}[htb]
  \begin{minipage}[t]{0.45\linewidth}%
    \centering
    \includegraphics[width=\linewidth]{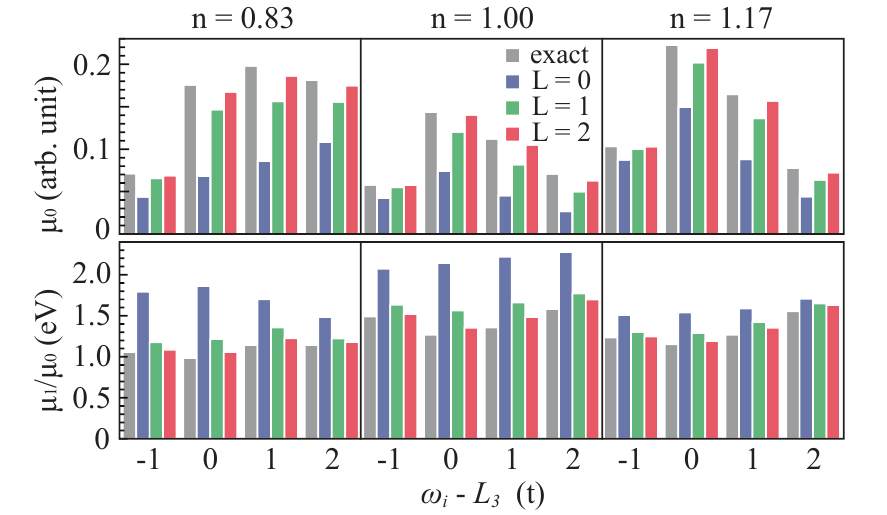}
    \caption{\label{Sfig:moment1dxx} Comparison of the zeroth moment $\mu_0$ and normalized first moment $\mu_1/\mu_0$ of the exact and approximate cross sections for 1D Hubbard model in the parallel-polarization for different $\omega_i$.}
  \end{minipage}\qquad
  \begin{minipage}[t]{0.45\linewidth}%
    \centering
    \includegraphics[width=\linewidth]{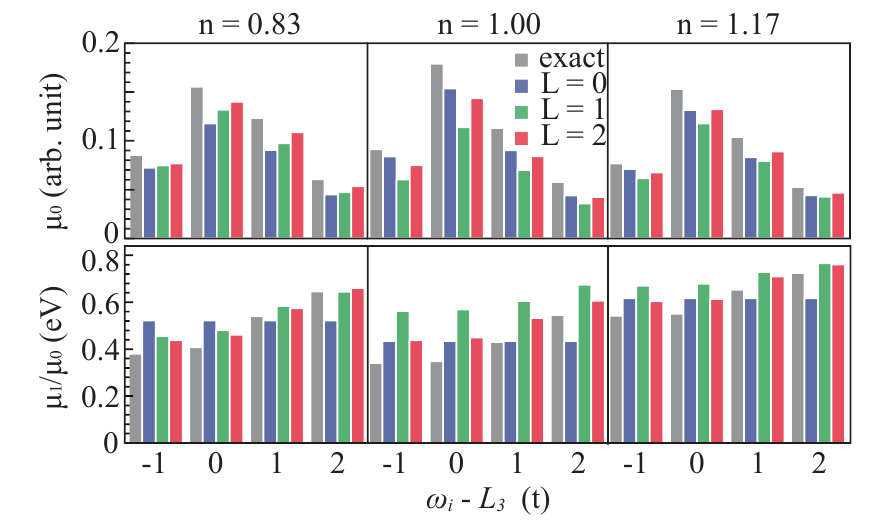}
    \caption{\label{Sfig:moment1dxy} Same as Fig.~\ref{Sfig:moment1dxx}, for the cross-polarization for different $\omega_i$.}
  \end{minipage}
  \end{figure}

\section{Range $l$ dependence of expansion coefficients $\alpha$ and exponential rate of convergence}

In Fig. 3 of the main text we show the expansion coefficient of $S_{z,0}$ and $j_{s,1}$. Here we show the full spectra for $l\leq 5$. Figs. 3(b) and 3(c) plot the corresponding values of $\alpha$ at $\omega_i=L_3$.

\begin{figure}[htp]
  \includegraphics[width=0.75\linewidth]{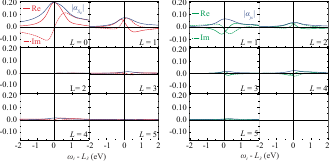}
  \caption{\label{Sfig:coeff_dist} Same as Fig.~3(a), for larger distances.}
\end{figure}

\section{Comparison of exact and approximated RIXS cross sections for the 2D Hubbard model}

\begin{figure}[htp]
  \begin{minipage}[t]{0.45\linewidth}
  \centering
  \includegraphics[width=\linewidth]{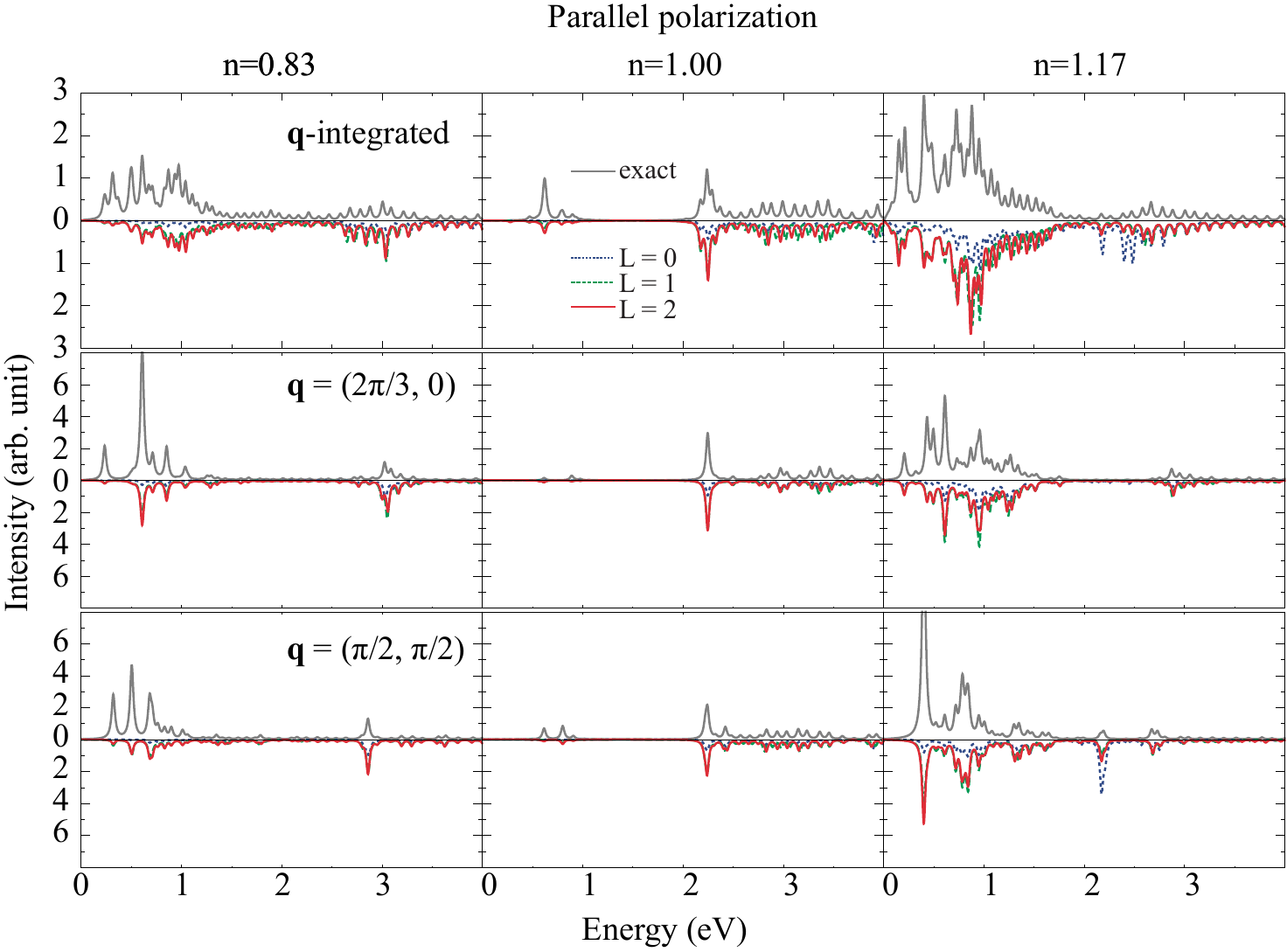}
  \caption{\label{Sfig:2dxx} Exact and approximate $\bm{q}$-integrated RIXS cross section for 2D Hubbard model in the parallel-polarization channel with $\omega_i=L_3$.}
\end{minipage}\qquad
\begin{minipage}[t]{0.45\linewidth}
  \centering
  \includegraphics[width=\linewidth]{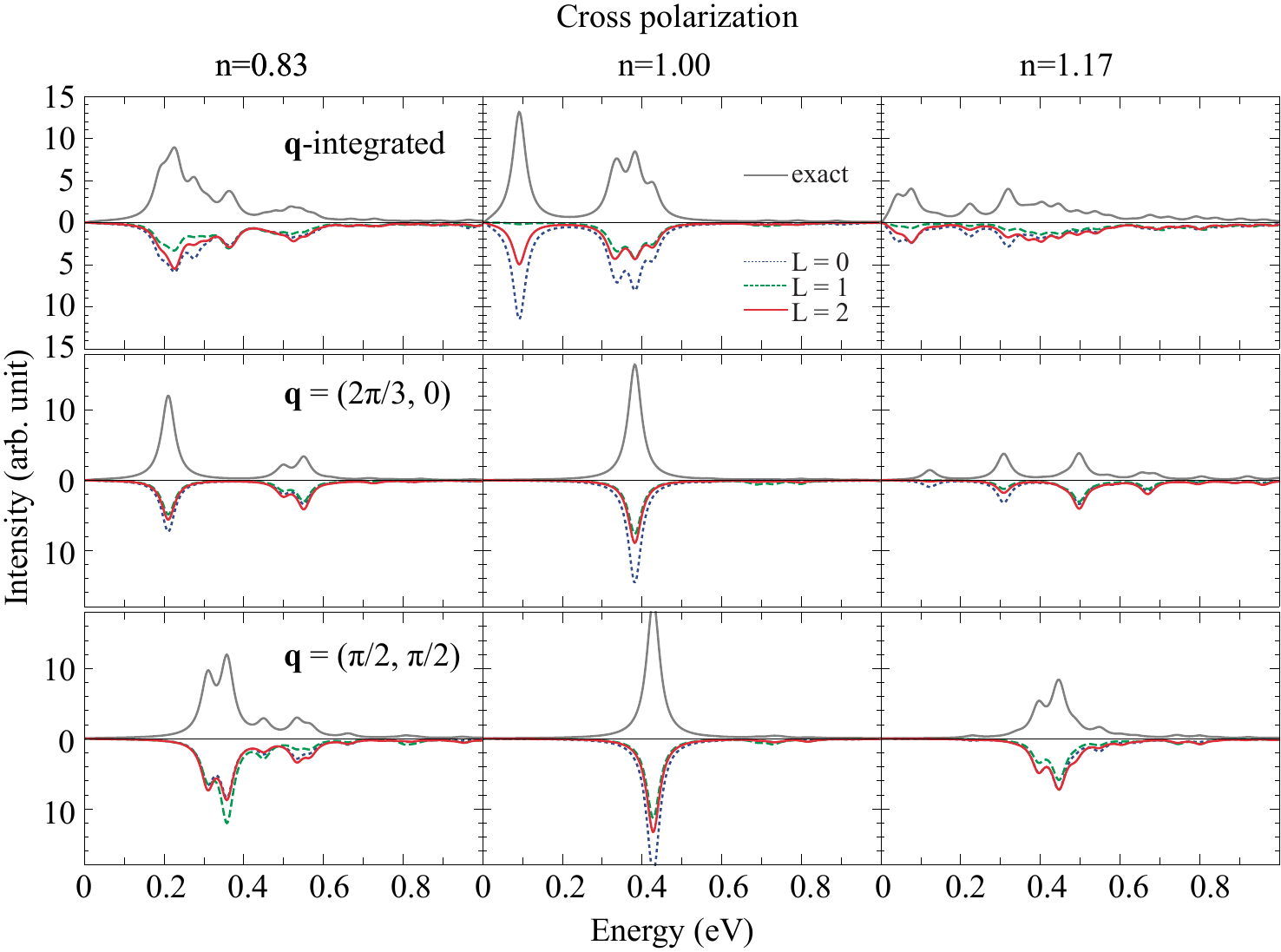}
  \caption{\label{Sfig:2dxy} Exact and approximate $\bm{q}$-integrated RIXS cross section for 2D Hubbard model in the cross-polarization channel with $\omega_i=L_3$.}
\end{minipage}
\end{figure}

Figure~\ref{Sfig:2dxx} and \ref{Sfig:2dxy} show the comparison of the exact and approximated results for the 2D Hubbard model.

